# Annotation of protein-coding genes in 49 diatom genomes from the Bacillariophyta clade

Natalia Nenasheva[1,^], Clara Pitzschel[1,^], Cynthia N. Webster[2], Alex Hart[2], Jill L. Wegrzyn[2], Mia M. Bengtsson[3], Katharina J. Hoff[1]

1. University of Greifswald, Institute of Mathematics and Computer Science and Center for Functional Genomics of Microbes, Walther-Rathenau-Str. 47, 17489 Greifswald, Germany
2. University of Connecticut, Department of Ecology and Evolutionary Biology, Plant Computational Genomics Lab, 75 N. Eagleville Road, Unit 3043 Storrs, CT 06269-3043, USA
3. University of Greifswald, Institute of Microbiology, Felix-Hausdorff-Straße 8, 17489 Greifswald, Germany

^) The authors contributed equally to this study corresponding author: Katharina J. Hoff ([katharina.hoff@uni-greifswald.de](katharina.hoff@uni-greifswald.de))

# Abstract

Diatoms, a major group of microalgae, play a critical role in global carbon cycling and primary production. Despite their ecological significance, comprehensive genomic resources for diatoms are limited. To address this, we have annotated previously unannotated genome assemblies of 49 diatom species.
Genome assemblies were obtained from NCBI Datasets and processed for repeat elements using RepeatModeler2 and RepeatMasker. For gene prediction, BRAKER2 was employed in the absence of transcriptomic data, while BRAKER3 was utilized when transcriptome short read data were available from the Sequence Read Archive. The quality of genome assemblies and predicted protein sets was evaluated using BUSCO, ensuring high-quality genomic resources. Functional annotation was performed using EnTAP, providing insights into the biological roles of the predicted proteins.
Our study enhances the genomic toolkit available for diatoms, facilitating future research in diatom biology, ecology, and evolution.

# Background & Summary

Diatoms are a diverse group of algae that significantly contribute to global carbon fixation and marine and freshwater ecosystem function [1]. In addition to their ecological role, their ability to tolerate and quickly acclimate to rapidly changing environmental conditions is remarkable [2]. These photosynthetic microalgae may capture and transmit $CO_2$ into diverse compounds, including lipids, omega-3 fatty acids, pigments, antioxidants, and polysaccharides (Sethi et al., 2020). They produce a variety of phytosterols, which offer possible health benefits such as cholesterol-lowering properties [3]. Diatoms can be cultivated indoors and outdoors, and their biomass productivity can be doubled in high-technology photobioreactors. A few selected species are used as model organisms in genetics and biochemistry research, while several taxa could be a bioprocess platform for biofuels [4].

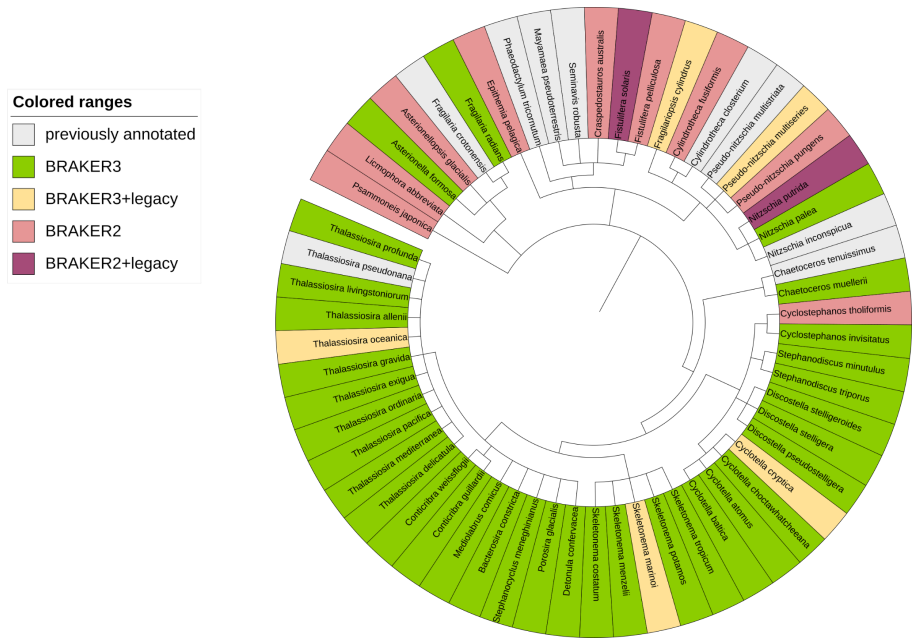

**Figure 1:** Taxonomy tree of selected Bacillariophyta genomes. This tree displays selected Bacillariophyta genome assemblies available from NCBI datasets between June 14th and 26th 2024. The tree was generated by PhyloT (https://phylot.biobyte.de/, August 21st 2024), visualized with iTol [55]. Representative genome assemblies with a previously existing annotation at NCBI are labeled in gray. Genomes labeled in yellow were here annotated using BRAKER3, including proteins from the same species that were already available for an older assembly or from PhycoCosm. Genomes labeled in blue were annotated with BRAKER2, including proteins from the same species that were already available for an older assembly or at PhycoCosm. Genomes labeled in green were annotated with BRAKER3 without legacy proteins, and genomes labeled in purple were annotated with BRAKER2 without legacy proteins. We excluded "uncultured" entries and those matching only two letters followed by a dot, e.g. "sp.".

Diatoms play a critical role in the global carbon cycle [4–6]. Through photosynthesis, diatoms convert carbon dioxide into organic carbon, forming the basis of marine food webs and assisting in the sequestration of carbon in ocean sediments [6]. Diatoms fix atmospheric carbon dioxide, accounting for around 20% of the world's primary production [7]. Their silica-based cell walls contribute to long-term carbon storage as they cause diatom cells to sink and settle on the ocean floor or the bottom of lakes and rivers. This process may be especially important during diatom blooms, which characterize temperate ocean margin zones and freshwater bodies in the spring. Various environmental factors in interactions with marine ecosystems affect the onset and progression of blooms, such as temperature, light intensity, and fluctuations of nutrients [8,9].

Interaction and coexistence with bacterial communities are an integral part of the life of diatom algae. They also form consortia and heterogeneous cohorts building networks of numerous cell-to-cell interactions for e.g. nutrient exchange. In this mutually beneficial deal, bacteria contribute by assimilating nutrients from the water and sequester minerals released by diatoms efficiently. Further, bacteria supply nutrients that diatoms are not able to produce themselves, for example, vitamins and fixed nitrogen [10]. Additionally, diatom blooms influence bacterial communities, showcasing their interconnectedness in marine ecosystems (e.g. [11–13]). At the same time, bacteria impact the dynamics of diatom growth [14]. The ecological roles of diatoms and their interaction with other organisms are now better-understood thanks to molecular techniques, which have provided new insights into cell death, silicon metabolism, environmental sensing, and community-level interactions [15].

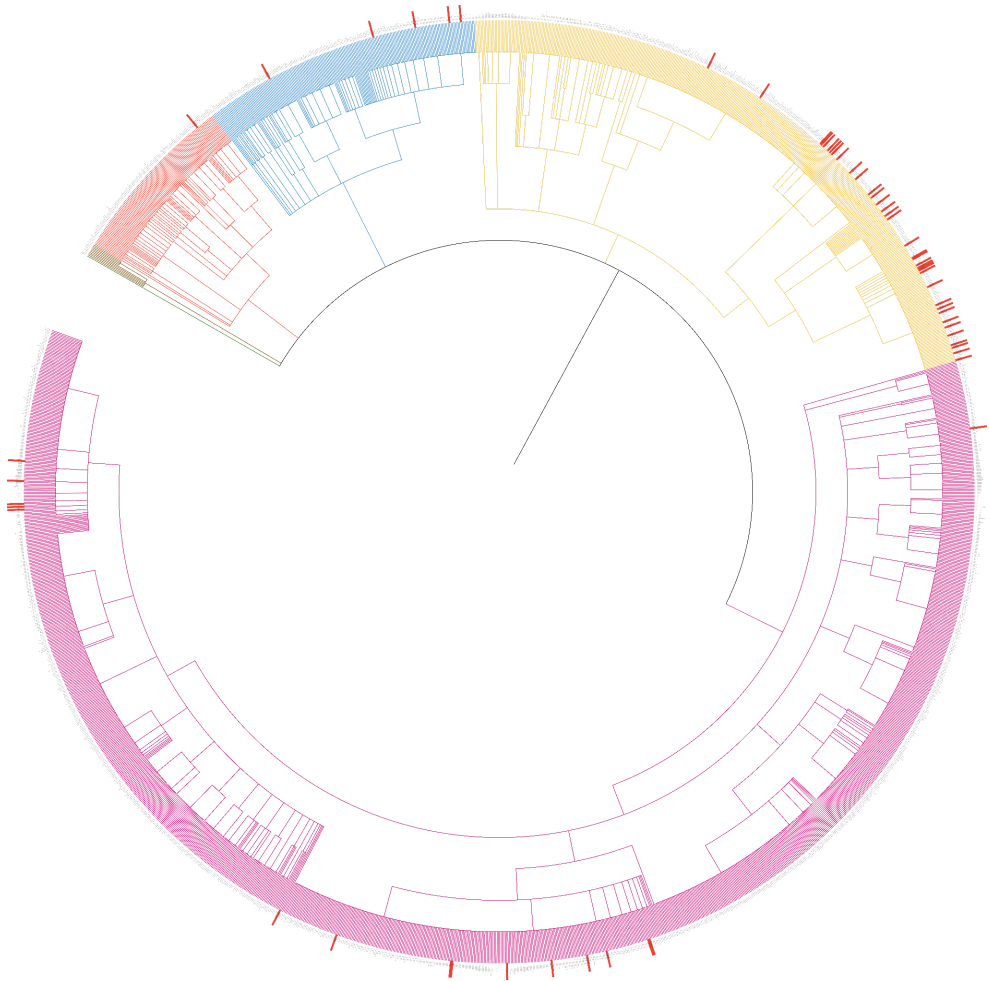

**Figure 2:** Taxonomic tree of all Bacillariophyta. Data was retrieved from NCBI Taxonomy via PhyloT, which automatically excluded many metagenomic isolates. The red outer leafs are available representative genomes that have been annotated for protein coding genes (both the previously existing 9 species and the newly annotated 49 species are labelled in red). Pink branches: Bacillariaphyceae; yellow: Coscinodiscophycaea; blue: Fragilariophycaea; brown: Bacillariophyta incertae sedis; green: Unclassified Bacillariophyta.

However, despite the frequency and importance of diatoms in the ecosystem, complete genetic resources for diatoms are scarce. When starting this study, we found 89 Bacillariophyta genome assemblies at National Center for Biotechnology Information (NCBI) Datasets (https://www.ncbi.nlm.nih.gov/datasets/, April 1st, 2024, see **Supplementary Table S1**). Of these, 66 were flagged as "representative genomes". In total, 13 of these genome assemblies had an annotation of protein coding genes, but only seven of the genome assemblies flagged as "representative genomes" had such an annotation. This means for four of the annotated assemblies, a younger and better but yet unannotated genome assembly existed (but the assembly of *Thalassiosira pseudonana* was not flagged as representative, had been annotated, and no alternative representative genome assembly was available). For three species available at the NCBI, we found an annotation of protein coding genes in PhycoCosm [16] but not at the NCBI. Knowledge about the protein coding genes is essential to fully exploit genome sequences [17], and thus we made it our mission to annotate previously unannotated genome assemblies of the Bacillariophyta.

| Species name | Literature Reference(s) | Data Reference(s) | Species name | Literature Reference(s) | Data Reference(s) |
|---|---|---|---|---|---|
| *Asterionella formosa* | unknown | 56,57 | *Nitzschia inconspicua* | 138 | 139 |
| *Asterionellopsis glacialis* | 58 | 59 | *Nitzschia palea* | unknown | 140–146 |
| *Bacterosira constricta* | 60 | 61,62 | *Nitzschia putrida* | 147 | 148 |
| *Chaetoceros muellerii* | 63 | 64–70 | *Phaeodactylum tricornutum* | 149 | 150 |
| *Chaetocerus tenuissimus* | 71 | 72 | *Porosira glacialis* | 60 | 151,152 |
| *Conticribra guillardii* | 60 | 73,74 | *Psammoneis japonica* | unknown | 153 |
| *Conticribra weissflogii* | 60 | 75–77 | *Pseudo-nitzschia multistrata* | 154,155 | 156 |
| *Craspedostauros australis* | 224 | 78 | *Pseudo-nitzschia multiseries* | unknown | 157–164 |
| *Cyclostephanos invisitatus* | 60 | 79,80 | *Pseudo-nitzschia pungens* | unknown | 165 |
| *Cyclostephanos tholiformis* | 60 | 81 | *Seminavis robusta* | 166 | 167 |
| *Cyclotella atomus* | 60 | 82–85 | *Skeletonema costatum* | 168 | 169–175 |
| *Cyclotella baltica* | 60 | 86,87 | *Skeletonema marinoi* | 176 | 177–183 |
| *Cyclotella choctawhatcheeana* | 60 | 88,89 | *Skeletonema menzelii* | 60 | 184,185 |
| *Cyclotella cryptica* | 60 | 90–95 | *Skeletonema potamos* | 60 | 186,187 |
| *Cylindrotheca closterium* | unknown | 96 | *Skeletonema tropicum* | 60 | 188,189 |
| *Cylindrotheca fusiformis* | 63 | 97 | *Stephanocyclus meneghinianus* | 60 | 190,191 |
| *Detonula confervacea* | 60 | 98–100 | *Stephanodiscus minutulus* | 60 | 192,193 |
| *Discostella pseudostelligera* | 60 | 101–103 | *Stephanodiscus triporus* | 60 | 194,195 |
| *Discostella stelligera* | 60 | 104,105 | *Thalassiosira allenii* | 60 | 196,197 |
| *Discostella stelligeroides* | 60 | 106,107 | *Thalassiosira delicatula* | 60 | 198–200 |
| *Epithemia pelagica* | unknown | 108 | *Thalassiosira exigua* | 60 | 201,202 |
| *Fistulifera pelliculosa* | 109 | 110 | *Thalassiosira gravida* | 60 | 203,204 |
| *Fistulifera solaris* | 109 | 111,112 | *Thalassiosira livingstoniorum* | 60 | 205 |
| *Fragilaria crotonensis* | 113,114 | 115 | *Thalassiosira mediterranea* | 60 | 206,207 |
| *Fragilaria radians* | 116 | 117–122 | *Thalassiosira oceanica* | 63 | 208–214 |
| *Fragilariopsis cylindrus* | 123 | 124–131 | *Thalassiosira ordinaria* | 60 | 215,216 |
| *Licmophora abbreviata* | unknown | 132 | *Thalassiosira pacifica* | 60 | 217–219 |
| *Mayamaea pseudoterrestris* | 133 | 134 | *Thalassioria pseudonana* | 220 | 221 |
| *Mediolabrus comicus* | 60 | 135–137 | *Thalassiosira profunda* | 60 | 222,223 |

**Table 1**: References for sequence data used in this study, either for genome annotation or for comparison to annotations of previously existing genome annotations. For some species, we were unable to retrieve an article reference (indicated with "unknown"). Data references are provided both for genome assemblies and transcriptome data.

Initially, we set out to annotate the genome assemblies of all Bacillariophyta that did not have an annotation of protein coding genes, or where a younger and better representative genome has been made available without annotation. Looking at redundancy (sometimes more than one genome assembly for the same species is available), we selected one assembly from each species. However, we decided later to exclude 10 genome assemblies (see **Supplementary Table S2**), either due to technical problems during download or annotation, or due to data quality. We ended up successfully annotating 49 Bacillariophyta genome assemblies (references to the original sequence data publications are listed in **Table 1**, genome assembly details are given in **Supplementary Table S3**, a taxonomic tree is shown in **Fig. 1**).

With this study, we present the annotation data of protein coding genes for 49 Bacillariophyta genome assemblies that were previously stored as unannotated at NCBI Datasets. Combined with the previously existing annotations, this now makes a total of 58 Bacillariophyta genome annotations accessible for further studies (**Fig. 2** visualizes how these 58 species cover the taxonomic tree of Bacillariophyta). Together, these data can be applied to various scientific problems and help researchers better understand many of the processes in diatom algae.

# Methods

The genome annotations presented here were generated using publicly available genome, transcriptome, and protein data. Data analysis was performed in three steps: (1) data preparation, (2) structural genome annotation, and (3) functional genome annotation. Steps 1 and 2 were executed using a semi-automated and reproducible Snakemake workflow [18] that is publicly available at https://github.com/KatharinaHoff/braker-snake (August 30th, 2024). Singularity [19] was employed to manage software dependencies. Step 3 was performed manually. All software version numbers are listed in **Supplementary Table S4**.

## Data preparation

In short, we used the NCBI Datasets tool to retrieve Bacillariophyta genome assembly information from the NCBI database. Assembly information was filtered to exclude 'uncultured' samples and species names ending in 'sp.' If multiple assemblies were available for the same species, we prioritized the 'representative' assembly, or, if unavailable, the assembly with the largest N50. Genomes with fewer than or equal to 1,000 annotated proteins were selected as candidates for further annotation. This threshold was set to include genome assemblies for annotation that have only a protein coding gene annotation for organelle genomes. For each candidate genome, we checked if an older assembly had existing protein-coding gene annotations (referred to as 'legacy proteins') and stored this information. All genome assemblies and any associated legacy proteins were downloaded using the datasets tool.

The workflow automatically retrieves the appropriate OrthoDB v11 partition [20] for the specified taxon from https://bioinf.uni-greifswald.de/bioinf/partitioned_odb11/. For Bacillariophyta, this corresponds to the Stramenopiles partition, which we combined with the Viridiplantae partition to ensure a larger sequence set.

For species lacking genome annotations, RNA-seq data availability was verified using the Biopython/Entrez API to query the Sequence Read Archive [21]. Up to six Illumina paired-end libraries were selected (the top six entries from the Entrez results), and downloaded using fasterq-dump (https://trace.ncbi.nlm.nih.gov/Traces/sra/sra.cgi?view=software, accessed August 21st, 2024). RNA-seq data were aligned to the genome using HISAT2 [22]. Co-culture libraries were not excluded, as they often provide critical data for diatoms, but libraries with an alignment rate below 20% were discarded. The resulting SAM files were converted to BAM, merged if multiple files existed, sorted, and indexed using SAMtools [23].

Before proceeding with automated annotation, we manually queried the PhycoCosm portal (Joint Genome Institute) for existing protein-coding gene annotations for species in our dataset. For *Cyclotella cryptica*, *Nitzschia putrida* and *Pseudo-nitzschia multiseries*, we downloaded available protein sequences and included them as 'legacy proteins' in the BRAKER annotation process.

The final output of this data preparation phase was a CSV file that specifies the input files required for the subsequent annotation workflow for each species.

## Structural genome annotation

Each selected genome assembly was processed individually using a consistent pipeline. First, RepeatModeler2 [24] was used to construct a species-specific repeat library, followed by RepeatMasker (http://www.repeatmasker.org, accessed August 21st, 2024) to soft mask the repeats in the genome. Depending on the availability of extrinsic data, either BRAKER2 or BRAKER3 [25,26] was employed to predict protein-coding gene structures from the soft-masked genome.

Protein evidence was always used during annotation. For many genomes, the combined Stramenopiles/Viridiplantae protein partition was used as input. Additionally, legacy proteins were incorporated when available. In cases where RNA-seq data were absent, BRAKER2 was run with an option to enrich the predicted gene set using BUSCOs from the Stramenopiles_odb10 dataset [27], enhanced with compleasm [28]. BRAKER2 first uses GeneMark-EP+ [29], which self-trains GeneMark-ES [30,31] to identify seed gene sequences. These sequences are then compared to the protein database using DIAMOND [32], followed by accurate spliced alignment with Spaln2 [33]. GeneMark-EP+ generates an intermediate gene set based on protein evidence, which is refined using AUGUSTUS [34,35]. TSEBRA [36] then combines and filters the predictions using protein evidence and BUSCOs as guides [37].

When RNA-seq alignments were available, BRAKER3 was used. This workflow employed GeneMark-ETP [38], which processes RNA-seq alignments using StringTie2 [39] to assemble transcripts. GeneMarkS-T [40] then screens the assembled transcripts for potential genes. DIAMOND and GeneMark-EP+'s protein evidence pipeline were used to filter the genes, and GeneMark-ETP also performed ab initio gene predictions based on self-training. AUGUSTUS was again trained on a reliable subset of predicted genes, and the final gene set was merged using TSEBRA.

Not all BRAKER jobs completed successfully; assemblies affected by these failures were excluded from further analysis (see **Table 1**).

For quality control, we ran BUSCO with the Stramenopiles_odb10 dataset on both the genome assemblies and the predicted protein sequences. Genomes were excluded if there was a significant discrepancy between BUSCO completeness scores at the genome level and the predicted protein level. For example, despite a 95% BUSCO completeness score at the genome level, *Pseudo-nitzschia delicatissima* achieved only 72% completeness at the annotation level and was excluded (see **Fig. 3**). Additionally, *Thalassiosira sundarbana* was excluded due to low genome BUSCO completeness (15%) and contamination in the database. *Epithemia catenata* was also excluded due to low genome BUSCO completeness (56%).

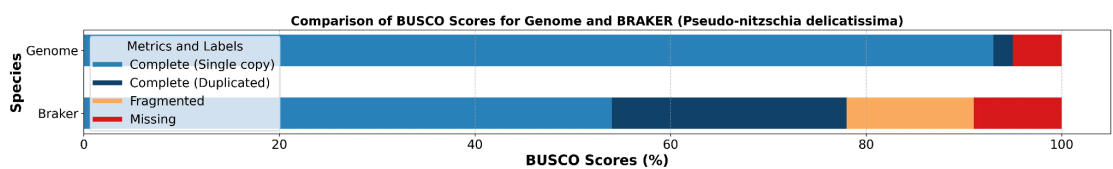

**Figure 3:** BUSCO scores of *Pseudo-nitzschia delicatissima*. We decided to exclude this species from further analysis because of the discrepancy of BUSCO scores between genome and protein level.

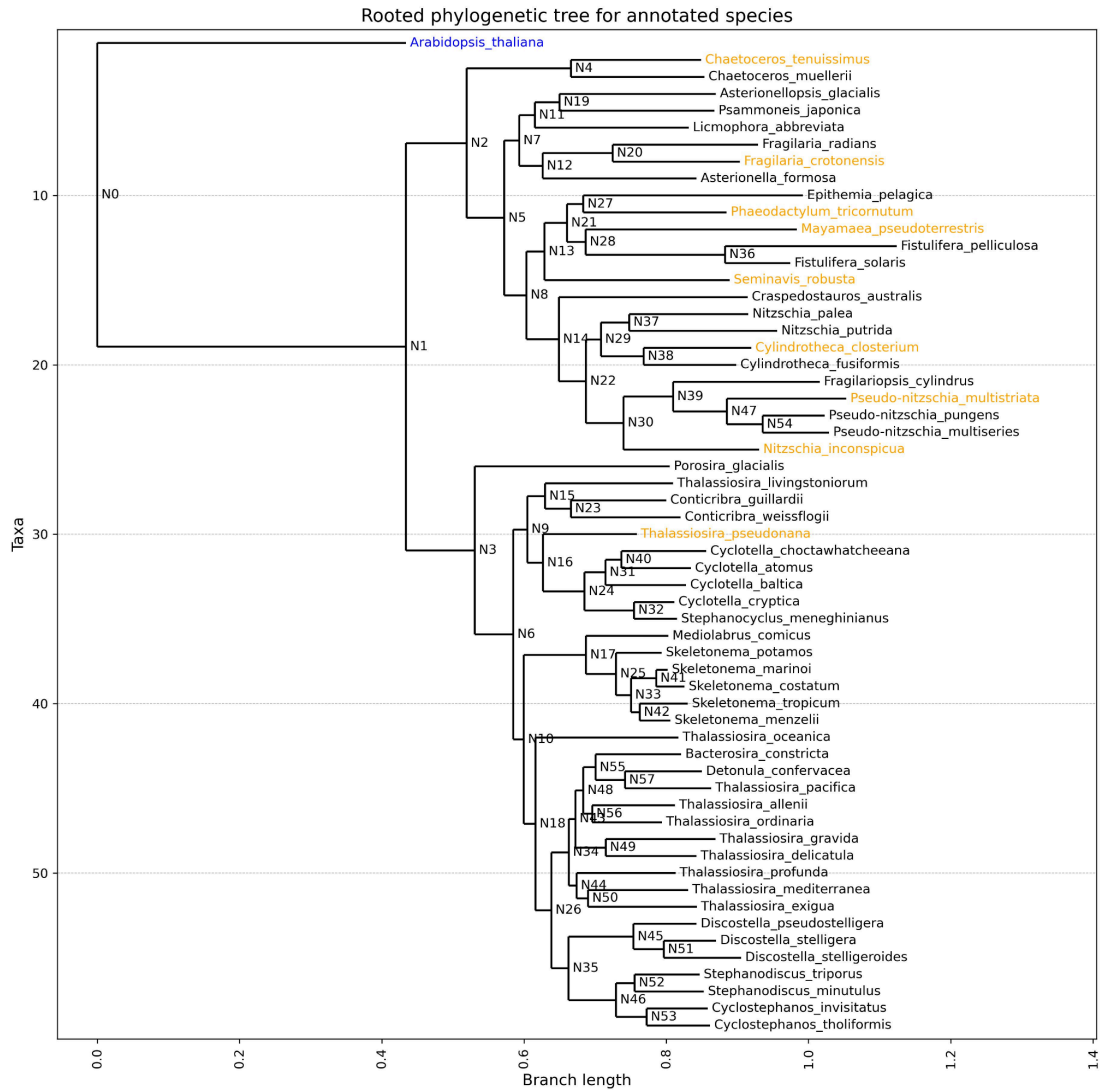

**Figure 4**: Rooted phylogenetic tree of Bacillariophyta with an annotation of protein coding genes. The labelled nodes (N) that describe the relationships between species on the basis of orthogroups. The model species *Arabidopsis thaliana* (blue) serves as an outgroup. The previously annotated species (*C. tenuissimus, C. closterium, F. crotonensis, M. pseudoterrestris, N. inconspicua, P. tricornutum, Pseudo-nitzschia multistriata, S. robusta,* and *T. pseudonana*) are emphasised by the orange colour. Species plotted with the black color were annotated in this study.

## Functional gene annotation

The EnTAP functional annotation software (v 1.3.0) was employed to provide functional descriptors and identify potential contaminants for the predicted proteins (Hart et al., 2020). EnTAP was configured with two curated databases, NCBI's RefSeq Protein [41] and UniProtKB/Swiss-Prot [42], for similarity searches, utilizing a 50% target and query coverage minimum, and a DIAMOND E-value threshold of 0.00001. An optimal alignment was selected for each protein query based on phylogenetic relevance, informativeness, and standard alignment quality metrics. Additionally, EnTAP performed independent searches against the EggNOG database [43] using the EggNOG-mapper toolbox [44]. The resulting gene family assignments, along with high-quality similarity search alignments, facilitated the subsequent connections to Gene Ontology terms [45,46], protein domains from Pfam [47], and pathway associations via KEGG [48].

# Gene set filtering

Descriptive statistics of the raw BRAKER output (see **Table 2**) and the EnTAP annotation rate (see **Supplementary Table S5**) suggested that BRAKER overpredicted single-exon genes in some cases. This issue has previously been reported in land plant annotations [49].

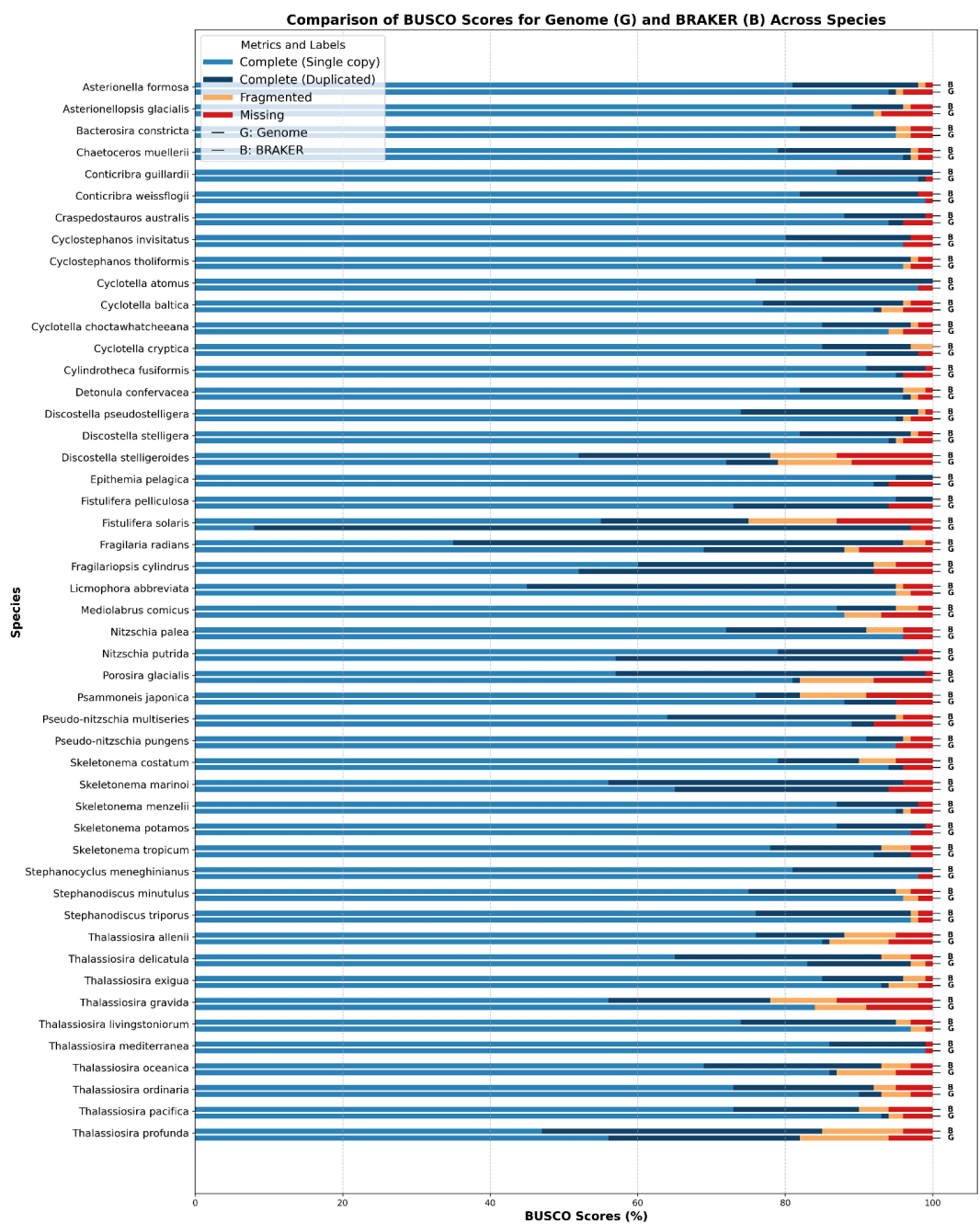

**Figure 5**: BUSCO results of genomes and protein sets. This plot demonstrates the quality of genome assemblies (G = Genome) and predicted protein sets (B = BRAKER) across all here annotated species; species ordered alphabetically. The categories Complete (Single copy or duplicated), Fragmented, or Missing BUSCOs are shown. Note that the duplicated predicted proteins are in part caused by the presence of alternative splicing isoforms in the BRAKER gene sets.

| Species name | #Genes | #Tx | Mono:Mult | Median Mult | Max Mult | Species name | #Genes | #Tx | Mono:Mult | Median Mult | Max Mult |
|---|---|---|---|---|---|---|---|---|---|---|---|
| *Asterionella formosa* | 17643 | 19690 | 2.44 | 2 | 15 | *Nitzschia palea* | 16112 | 19154 | 1.75 | 2 | 16 |
| *Asterionellopsis glacialis* | 19134 | 20492 | 1.59 | 2 | 16 | *Nitzschia putrida* | 20530 | 22856 | 1.3 | 2 | 13 |
| *Bacterosira constricta* | 22749 | 25062 | 1.98 | 2 | 30 | *Porosira glacialis* | 48999 | 49187 | 6.65 | 2 | 31 |
| *Chaetoceros muellerii* | 13586 | 15972 | 1.85 | 2 | 26 | *Psammoneis japonica* | 22827 | 24371 | 1.17 | 2 | 19 |
| *Conticribra guillardii* | 17169 | 19531 | 1.28 | 3 | 23 | *Pseudo-nitzschia multiseries* | 41854 | 44440 | 2.93 | 2 | 24 |
| *Conticribra weissflogii* | 15985 | 17945 | 1.29 | 3 | 20 | *Pseudo-nitzschia pungens* | 17930 | 19302 | 1.36 | 2 | 25 |
| *Craspedostauros australis* | 19297 | 21075 | 1.58 | 2 | 14 | *Skeletonema costatum* | 20649 | 22430 | 1.56 | 2 | 16 |
| *Cyclostephanos invisitatus* | 13354 | 16070 | 1.18 | 3 | 23 | *Skeletonema marinoi* | 22888 | 26524 | 2.46 | 2 | 17 |
| *Cyclostephanos tholiformis* | 19600 | 20902 | 0.8 | 3 | 34 | *Skeletonema menzelii* | 14214 | 17490 | 1.92 | 2 | 12 |
| *Cyclotella atomus* | 17063 | 19759 | 1.02 | 3 | 29 | *Skeletonema potamos* | 14321 | 17682 | 1.98 | 2 | 14 |
| *Cyclotella baltica* | 20650 | 23038 | 1.28 | 3 | 54 | *Skeletonema tropicum* | 23816 | 27647 | 2.77 | 2 | 14 |
| *Cyclotella choctawhatcheeana* | 16876 | 19331 | 1.01 | 3 | 50 | *Stephanocyclus meneghinianus* | 17489 | 20102 | 1.29 | 3 | 27 |
| *Cyclotella cryptica* | 27018 | 30077 | 1.83 | 3 | 55 | *Stephanodiscus minutulus* | 13530 | 15516 | 1.31 | 2 | 27 |
| *Cylindrotheca fusiformis* | 18449 | 19938 | 1.28 | 2 | 14 | *Stephanodiscus triporus* | 14465 | 16751 | 1.38 | 2 | 39 |
| *Detonula confervacea* | 25217 | 27369 | 2.29 | 2 | 23 | *Thalassiosira allenii* | 25993 | 27929 | 2.08 | 2 | 34 |
| *Discostella pseudostelligera* | 10867 | 14109 | 0.93 | 3 | 28 | *Thalassiosira delicatula* | 34282 | 36115 | 3.56 | 2 | 38 |
| *Discostella stelligera* | 14843 | 17755 | 1.36 | 3 | 40 | *Thalassiosira exigua* | 25701 | 28025 | 1.99 | 3 | 40 |
| *Discostella stelligeroides* | 16328 | 18945 | 1.36 | 2 | 40 | *Thalassiosira gravida* | 26431 | 27989 | 1.89 | 2 | 39 |
| *Epithemia pelagica* | 18983 | 21013 | 1.71 | 2 | 18 | *Thalassiosira livingstoniorum* | 44697 | 47357 | 3.79 | 3 | 38 |
| *Fistulifera pelliculosa* | 17328 | 21235 | 0.32 | 3 | 37 | *Thalassiosira mediterranea* | 19055 | 21492 | 1.38 | 3 | 49 |
| *Fistulifera solaris* | 26808 | 31306 | 0.8 | 2 | 18 | *Thalassiosira oceanica* | 26307 | 29507 | 1.14 | 3 | 42 |
| *Fragilaria radians* | 21031 | 23879 | 2.45 | 2 | 13 | *Thalassiosira ordinaria* | 28461 | 31353 | 1.8 | 3 | 32 |
| *Fragilariopsis cylindrus* | 21946 | 25304 | 1.74 | 2 | 24 | *Thalassiosira pacifica* | 29416 | 31996 | 2.54 | 2 | 29 |
| *Licmophora abbreviata* | 15275 | 16375 | 1.25 | 2 | 15 | *Thalassiosira profunda* | 25451 | 29337 | 1.66 | 2 | 19 |
| *Mediolabrus comicus* | 15450 | 18098 | 2.26 | 2 | 14 | | | | | | |

**Table 2:** Statistics on the raw BRAKER gene sets. #Genes, number of genes; #Tx, number of transcripts; Mono:Mult, mono-exon to multi-exon ratio; Median Mult, median number of exons in multi-exon genes; Max Mult, largest number of exons in multi-exon genes. Note that the gene sets described here are an intermediate result prior the final filtering.

To address this and filter out potential false positive single-exon gene predictions—while retaining gene models that may be of scientific interest—we applied the following filtering approach: We discarded single-exon gene models that lacked a functional annotation by EnTAP, did not have a significant hit in a DIAMOND search against the NCBI RefSeq non-redundant proteins (NR) database (February 2nd, 2024), and were not part of an orthologous group spanning more than one diatom species. For the final step, we conducted an initial OrthoFinder [50] analysis using the predicted proteomes from 49 diatom genomes.

## File processing

In order to prepare GFF3 files for upload to NCBI genomes, the filtered BRAKER output files were decorated with product names and notes according to EnTAP results (command lines at https://github.com/Gaius-Augustus/Diatom_annotation_scripts ).

## Orthogroup analysis

We used OrthoFinder to identify orthologous gene groups across species by performing an all-versus-all comparison of protein sequences (using the longest isoform of each gene). Based on sequence similarities, genes were grouped into orthogroups, which represent sets of genes descended from a common ancestor. To ensure the reliability of the phylogenetic tree, we included well-characterized model species *Arabidopsis thaliana* (annotation used in [25], available at https://github.com/gatech-genemark/EukSpecies-BRAKER2/blob/master/Arabidopsis_thaliana/annot/annot.gtf.gz) and nine publicly available annotations (see **Table 3**).

| Species name | Genome assembly accession number | Annotation source | #Genes | #Tx | Mono:Mult | Median Mult | Max Mult |
|---|---|---|---|---|---|---|---|
| Chaetoceros tenuissimus | GCA_021927905.1 | DDBJ | 18397 | 18670 | 1.38 | 2 | 96 |
| Cylindrotheca closterium | GCA_933822405.4 | EMBL | 24371 | 24633 | 1.53 | 2 | 18 |
| Fragilaria crotonensis | GCA_022925895.1 | Genbank | 26015 | 26015 | 2.14 | 2 | 15 |
| Mayamaea pseudoterrestris | GCA_027923505.1 | DDBJ | 11017 | 11017 | 1.04 | 2 | 20 |
| Nitzschia inconspicua | GCA_019154785.2 | Genbank | 38391 | 38393 | 1.37 | 2 | 14 |
| Phaeodactylum tricornutum | GCA_000150955.2 | Genbank | 10321 | 10339 | 1.19 | 2 | 13 |
| Pseudo-nitzschia multistriata | GCA_900660405.1 | EMBL | 11909 | 11952 | 1.46 | 2 | 18 |
| Seminavis robusta | GCA_903772945.1 | EMBL | 35858 | 35865 | 1.44 | 2 | 44 |
| Thalassiosira pseudonana | GCA_000149405.2 | Genbank | 11681 | 11686 | 0.66 | 3 | 77 |

**Table 3**: Descriptive statistics of the previously existing annotations of protein coding genes in representative genome assemblies at NCBI. #Genes, number of genes; #Tx, number of transcripts; Mono:Mult, mono-exon to multi-exon ratio; Median Mult, median number of exons in multi-exon genes; Max Mult, largest number of exons in multi-exon genes.

The species tree was inferred using the Species Tree from All Genes (STAG) algorithm (Emms and Kelly, 2018), which constructs a consensus tree from gene trees. We included support values at internal nodes for robustness, and the tree was rooted using the Species Tree Root Inference from Duplication Events (STRIDE) algorithm (Emms and Kelly, 2017). This approach leveraged both orthogroup and gene duplication information to ensure a robust phylogenetic inference.

# Data Records

We provide both structural and functional annotations of protein-coding genes for 49 Bacillariophyta genome assemblies, available from Zenodo (DOI: 10.5281/zenodo.13745090, URL: https://zenodo.org/records/13767023). The final annotation set is here described in **Table 4.** Supplementary Tables are separately available from Zenodo (DOI: 10.5281/zenodo.13900512, URL: https://zenodo.org/records/13900512 ).

| Species name | #Genes | #Tx | Mono:Mult | Median Mult | Max Mult | Species name | #Genes | #Tx | Mono:Mult | Median Mult | Max Mult |
|---|---|---|---|---|---|---|---|---|---|---|---|
| Asterionella formosa | 15671 | 17562 | 2.07 | 2 | 15 | Nitzschia palea | 14901 | 17738 | 1.55 | 2 | 16 |
| Asterionellopsis glacialis | 17168 | 18417 | 1.34 | 2 | 16 | Nitzschia putrida | 19020 | 21186 | 1.13 | 2 | 13 |
| Bacterosira constricta | 20515 | 22787 | 1.71 | 2 | 30 | Porosira glacialis | 26854 | 27981 | 3.34 | 2 | 32 |
| Chaetoceros muellerii | 12260 | 14501 | 1.58 | 2 | 26 | Psammoneis japonica | 20769 | 22183 | 0.98 | 2 | 21 |
| Conticribra guillardii | 15702 | 18002 | 1.1 | 3 | 23 | Pseudo-nitzschia multiseries | 29082 | 34237 | 2 | 2 | 24 |
| Conticribra weissflogii | 14195 | 16094 | 1.05 | 3 | 31 | Pseudo-nitzschia pungens | 16606 | 17903 | 1.19 | 2 | 25 |
| Craspedostauros australis | 17212 | 18700 | 1.3 | 2 | 14 | Skeletonema costatum | 20346 | 22118 | 1.52 | 2 | 16 |
| Cyclostephanos invisitatus | 13041 | 15742 | 1.13 | 3 | 23 | Skeletonema marinoi | 22888 | 26524 | 2.46 | 2 | 17 |
| Cyclostephanos tholiformis | 18632 | 19916 | 0.71 | 3 | 34 | Skeletonema menzelii | 14043 | 17300 | 1.88 | 2 | 12 |
| Cyclotella atomus | 16269 | 18917 | 0.93 | 3 | 51 | Skeletonema potamos | 14068 | 17400 | 1.93 | 2 | 14 |
| Cyclotella baltica | 19068 | 21394 | 1.11 | 3 | 54 | Skeletonema tropicum | 22700 | 26385 | 2.6 | 2 | 14 |
| Cyclotella choctawhatcheeana | 16069 | 18466 | 0.92 | 3 | 50 | Stephanocyclus meneghinianus | 16500 | 18726 | 1.13 | 3 | 41 |
| Cyclotella cryptica | 24916 | 27872 | 1.62 | 3 | 55 | Stephanodiscus minutulus | 12748 | 14714 | 1.19 | 2 | 27 |
| Cylindrotheca fusiformis | 16771 | 18166 | 1.08 | 2 | 14 | Stephanodiscus triporus | 13961 | 16219 | 1.31 | 2 | 39 |
| Detonula confervacea | 22240 | 24347 | 1.92 | 2 | 23 | Thalassiosira allenii | 23169 | 25057 | 1.76 | 2 | 34 |
| Discostella pseudostelligera | 10694 | 13923 | 0.91 | 3 | 31 | Thalassiosira delicatula | 28894 | 30678 | 2.87 | 2 | 38 |
| Discostella stelligera | 14199 | 17104 | 1.27 | 3 | 40 | Thalassiosira exigua | 21764 | 23986 | 1.56 | 3 | 40 |
| Discostella stelligeroides | 15804 | 18403 | 1.3 | 2 | 40 | Thalassiosira gravida | 23308 | 24824 | 1.55 | 2 | 39 |
| Epithemia pelagica | 17109 | 20995 | 0.3 | 3 | 37 | Thalassiosira livingstoniorum | 34258 | 36738 | 2.71 | 2 | 38 |
| Fistulifera pelliculosa | 26636 | 31123 | 0.79 | 2 | 18 | Thalassiosira mediterranea | 17908 | 20287 | 1.24 | 3 | 49 |
| Fistulifera solaris | 18991 | 21642 | 2.13 | 2 | 13 | Thalassiosira oceanica | 25523 | 28710 | 1.08 | 3 | 42 |
| Fragilaria radians | 21576 | 24906 | 1.7 | 2 | 14 | Thalassiosira ordinaria | 25978 | 28794 | 1.57 | 3 | 32 |
| Fragilariopsis cylindrus | 14512 | 15570 | 1.14 | 2 | 15 | Thalassiosira pacifica | 25522 | 28029 | 2.1 | 2 | 29 |
| Licmophora abbreviata | 15671 | 17562 | 2.07 | 2 | 15 | Thalassiosira profunda | 24879 | 28765 | 1.61 | 2 | 19 |
| Mediolabrus comicus | 14744 | 17323 | 2.12 | 2 | 14 | | | | | | |

**Table 4:** Statistics on the final BRAKER gene sets. #Genes, number of genes; #Tx, number of transcripts; Mono:Mult, mono-exon to multi-exon ratio; Median Mult, median number of exons in multi-exon genes; Max Mult, largest number of exons in multi-exon genes.

## Technical Validation

We performed a genome annotation study focusing on 49 diatom species, aiming to create a robust genomic dataset that supports future research into diatom biology and evolution. To emphasize the need for our work, we plotted the distribution of existing Bacillariophyta genome annotations in the context of all known species within this taxon (**Fig. 2**). This analysis highlights the limited representation of annotated diatom species in current genomic resources. Our work significantly expands the number of annotated assemblies from 9 (or 15, including legacy assemblies) to 58, providing a valuable resource for diatom research.

Descriptive statistics for the gene structures of the newly annotated genomes are provided in **Table 4**. Previously annotated diatom genomes at NCBI contain between 10,321 and 38,391 protein-coding gene models (see **Table 3**). The gene numbers in the newly generated gene sets fall within this range. [49] recommend using the ratio of mono-exon to multi-exon genes as a quality measure for genome annotations, with a suggested ratio of 0.2 for land plants. In contrast, diatom genomes exhibit a higher proportion of single-exon genes, ranging from 0.66 to 2.14 (based on existing annotations; see **Table 3**). The BRAKER2 and BRAKER3 pipelines tend to overpredict single-exon genes, and we hypothesize that this phenomenon extends to diatom genomes as well. After applying our filtering approach, only five species - *Porosira glacialis* (3.34), *Skeletonema marinoi* (2.46), *Skeletonema tropicum* (2.6), *Thalassiosira delicatula* (2.87), and *Thalassiosira mediterranea* (2.71) - exceeded this range. These deviations are modest and may partly be attributed to selfish DNA elements, such as unmasked transposons and inserted retroviruses. The exon structure of the novel annotations aligns with previously annotated genomes in terms of the median number of exons per gene (2–3) and the largest number of exons per transcript (13–96) (see **Tables 4** and **3**).

Evaluating the quality of novel genome annotations is challenging. We used BUSCO to assess genome completeness at both the genome and protein levels, following Earth BioGenome Project guidelines [17]. BUSCO estimates the proportion of genes typically present as single copies within a clade. However, the stramenopiles_odb10 dataset applicable to diatoms is relatively small (100 marker genes). While BUSCO scores measure sensitivity within this limited dataset (**see Fig. 5**), a close agreement between genome- and protein-level scores suggests that the new annotations do not lack a significant portion of BUSCO genes detectable at the genome level. This is expected, as the stramenopiles_odb10 dataset was used as input for BRAKER. It is important to note that BUSCO does not correctly handle alternative transcripts, potentially skewing the number of duplicate BUSCOs in the protein set.

We also applied OMArk [51] to further assess the quality of protein-coding gene annotations. OMArk uses conserved homologous genes (HOGs) from the OMA database [52] and the OMAmer software for fast protein placement [53]. For Bacillariophyta, the relatively small Ochrophyta subset of 942 HOGs is applicable. While this is a limited number of marker genes, OMArk provides additional metrics, assessing contamination, consistency, and fragmentation. **Fig. 6** shows OMArk results for our newly annotated genomes, while **Fig. 7** displays results for previously available reference genomes. Unlike BUSCO, OMArk correctly handles alternative transcript isoforms, suggesting that the observed duplicates are likely real. Notably, we observed a high level of HOG completeness across most assemblies. However, *Thalassiosira profunda* and *Fistulifera solaris* showed a surprisingly high number of duplicate HOGs. For *T. profunda*, this is consistent with BUSCO scores at the genome level, indicating agreement between different metrics. In contrast, the source of duplicates in *F. solaris* remains unclear. We explored the genome assembly statistics (**Fig. 8**) but found no obvious explanation. Additionally, OMArk identified a significant level of contamination in the genome of *Licmophora abbreviata*, which had not been flagged as contaminated in public databases.

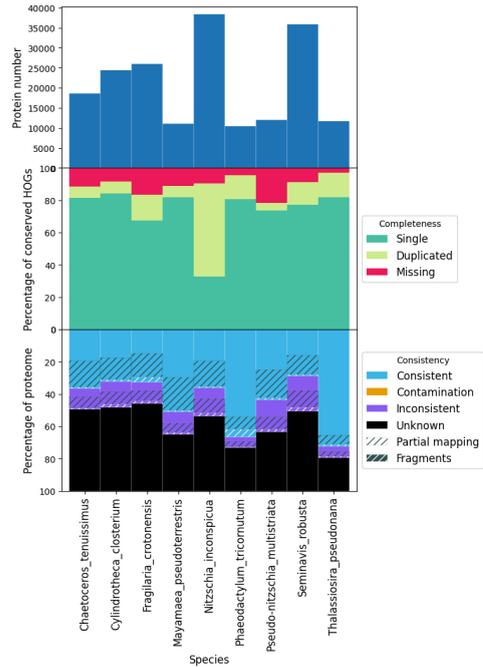

**Figure 7:** OMArk results of previously annotated Bacillariophyta reference genome assemblies. Since it was not straight-forward to extract alternative isoform nesting from the GFF3 files, we extracted the longest isoform for each locus with TSEBRA instead of generating an isoform information file for OMArk.

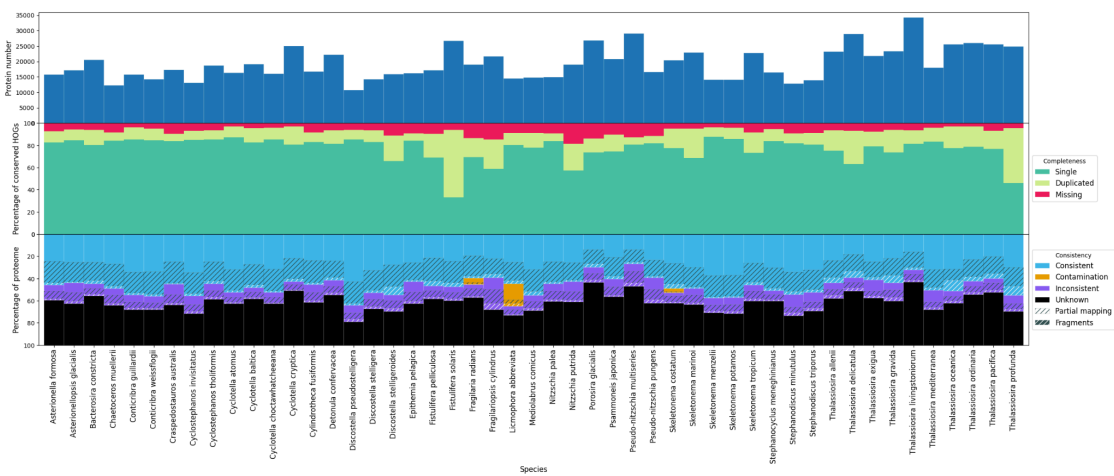

**Figure 6:** OMArk results of newly annotated Bacillariophyta genomes.

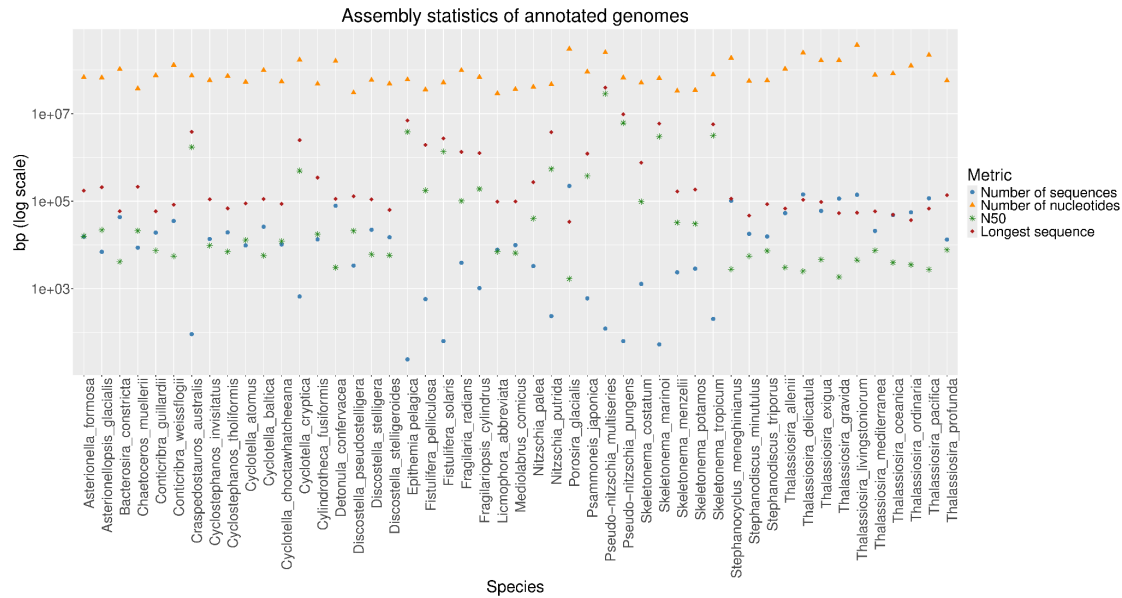

**Figure 8:** Assembly statistics of Bacillariophyta genome assemblies annotated with BRAKER. Note that the y-axis has a logarithmic scale.

To assess the quality of the resulting annotations and construct a phylogenetic tree, we analyzed the phylogenetic relationships between species using OrthoFinder. OrthoFinder identifies orthologous genes across species by determining the correspondence between genes and tracing their common ancestry. **Fig. 9** shows the proportion of genes assigned to orthogroups in this study. For most species, nearly 100% of transcripts were assigned to orthogroups, with only a small fraction remaining unassigned. It should be noted that OrthoFinder also constructs intra-species orthogroups, which consist of genes from a single species. Across all species, 97.9% of genes were assigned to these intra-species orthogroups.

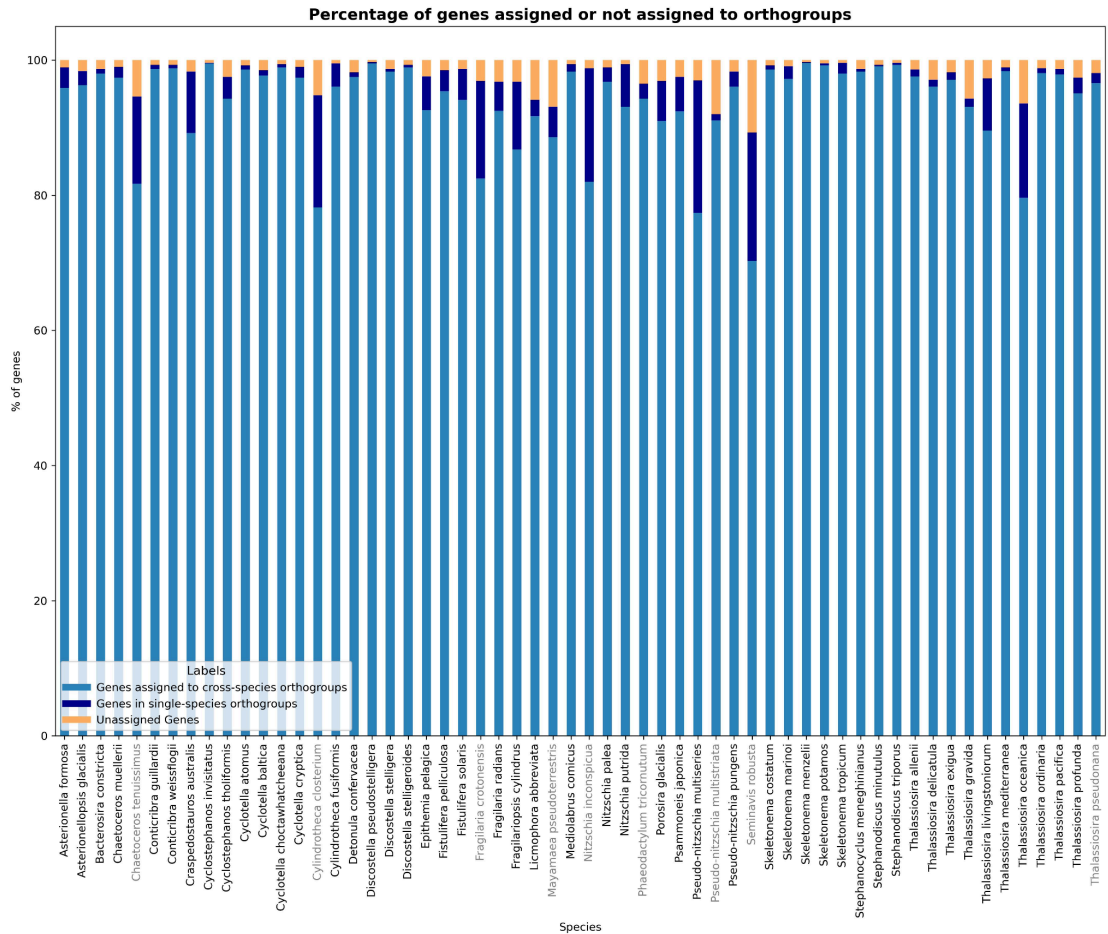

**Figure 9**: Assignment rate of proteins to orthogroups. The figure shows the percentage of genes assigned to cross-species orthogroups, to single-species orthogroups (paralog only groups), or not assigned to orthogroups across different species as a stacked barplot. The light blue bars represent the genes assigned to orthogroups, while the orange bars represent the unassigned genes. The dark blue bars represent the genes assigned to single-specie orthogroups. Previously annotated species are marked in gray.

In total, 953,148 transcripts (93.4% of the dataset) were assigned to inter-species orthogroups, emphasizing the significant degree of genetic overlap among the species included in this study. *Thalassiosira livingstoniorum* had the highest number of genes within inter-species orthogroups, with 30,692 genes, which also corresponds to the species with the largest number of annotated genes. This suggests that *T. livingstoniorum* has a more extensive set of orthologous genes, potentially indicating greater functional diversity.

The distribution of orthogroup sizes shows that most orthogroups are relatively small, with the majority falling within the 21–50 member range (see **Fig. 10**). Orthogroups with 11–20 members are also common, though less so than the 21–50 range. Large orthogroups, with more than 100 genes, are rare, and very few exceed 500 members. This pattern suggests that gene families in these species tend to be species-specific or functionally specialized, with a limited representation of highly conserved gene clusters across multiple species. The scarcity of large orthogroups may imply a divergence in evolutionary processes, leading to more functionally distinct gene groups.

OrthoFinder's phylogenetic analysis is based on the construction of gene trees, allowing for the classification of orthologous and paralogous relationships. Phylogenetic trees are particularly useful for identifying variable rates of sequence evolution (through branch lengths) and the order in which sequences diverged (tree topology). The resulting phylogenetic tree for Bacillariophyta genome assemblies, including both novel and previously annotated genomes from the International Nucleotide Sequence Database Collaboration (INSDC), is shown in **Fig. 4**.

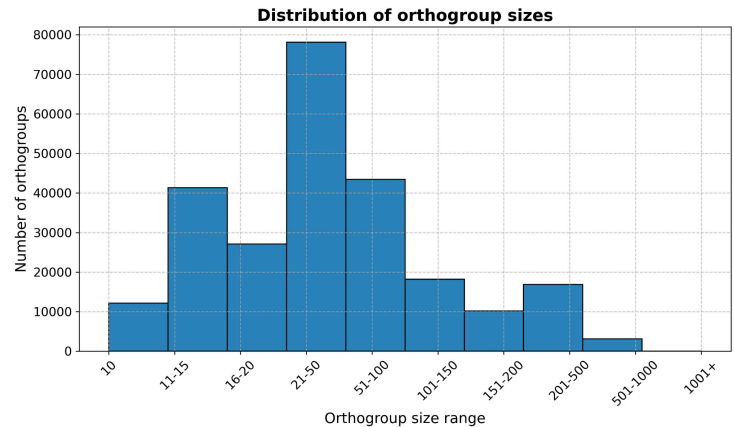

**Figure 10:** Distribution of Orthogroup Sizes.

While the PhycoCosm database includes additional annotated Bacillariophyta genomes, our workflow was specifically designed to rely on automatic querying of NCBI datasets for genome downloads. Therefore, we did not include PhycoCosm genomes in this study.

The novel annotations presented here will be valuable for studying interactions between diatoms and bacteria, particularly in the context of algal blooms that play a significant role in global carbon cycling. Given that methods for recovering full eukaryotic genomes from metagenomes are still developing, reference-based binning approaches, such as BlobTools [54] using DIAMOND, may provide a viable strategy, especially as databases like NCBI NR expand for this clade.

# Code Availability

The snakemake workflow used to generate this data set is freely available at https://github.com/KatharinaHoff/braker-snake . The postprocessing steps including custom scripts are freely available at https://github.com/Gaius-Augustus/Diatom_annotation_scripts . Container and software versions are listed in **Supplementary Table S4**.

# Acknowledgements

We thank Stepan Saenko for discussing the snakemake workflow for structural genome annotation. NN was funded by the German Research Foundation (DFG) in the framework of the research unit FOR2406 "Proteogenomics of Marine Polysaccharide Utilization (POMPU)" by a grant to KJH (HO 4545/4-3). CW and AH, and their software, were supported by a grant to JLW (NSF DBI 1943371).

## Author contributions

KJH and MMB conceptualized the study. CP and KJH designed and implemented the software. CP executed the snakemake workflow for structural genome annotation under supervision of KJH. CW and AH performed functional annotation under supervision of JLW. NN performed orthology analysis under supervision of KJH. NN wrote the first draft of the manuscript. All authors contributed to the manuscript in writing. All authors read and approved of the manuscript.

# Competing interests

No competing interests were disclosed.